# Features of the fluctuation –electromagnetic interaction between a small conducting particle and polarizable medium

G.V.Dedkov[1], A.A.Kyasov

Nanoscale Physics Group, Kabardino –Balkarian State University, Nalchik, 360004, Russian Federation

**Abstract. –** For the first time**,** new important features of the fluctuation electromagnetic interaction between a small conducting particle and a smooth surface of polarizable medium (both dielectric and metallic) are worked out. The particle is characterized by classical electric and magnetic polarizabilities. The temperature dependence and retardation effects are explicitly taken into account. The resulting interaction force between a metallic particle and the surface of metal proves to be determined to great extent by magnetic coupling and reveals specific dependences on distance, temperature, particle radius and material properties of contacting materials. Numerical estimations are given in the case of a Cu particle above a smooth Cu substrate at different particle radius and temperature of the system.

[1] Corresponding author E –mail: gv_dedkov@mail.ru

## 1.Introduction

This paper aims to show that conservative fluctuation electromagnetic force between a small conducting albeit nonmagnetic particle and the solid surface has new important features as compared with that one in the case of a dielectric particle or a ground state atom. In case of interest, in addition to the fluctuating electric moment, the particle has significant magnetic moment, which appreciably influences interaction. The fluctuating magnetic moment appears even on a resting nonmagnetic object due to stochastic Foucault currents being induced by non stationary or fluctuating external magnetic fields penetrating inside a volume of the object having non zero magnetic polarizability [1].

It is very worthwhile to get a better understanding of the fluctuation electromagnetic interaction because it fundamental and practical importance (for general discussion see Ref. [2]). Nowadays, a growing interest to interaction of atoms and nanoprobes with quantum electromagnetic field (EMF) close to and far from polarizable bodies with different material properties and temperature is being triggered by direct measurement of the Casimir forces [3],[4],[5], studies of their influence on micrometer –sized mechanical devices [6-8], dynamics of probing asperities in the scanning probe microscopy [5],[9],[10], nature of decoherence in ion traps due to external noise [11], etc.

Quite recently we have presented general formalism which allows solving a lot of problems on radiation heat exchange and conservative/dissipative forces acting on neutral atoms and nanoparticles moving nearby the solid surface or passing through the equilibrium background radiation [12],[13],[14],[15]. Within the dipole approximation of the theory, the starting expression for the particle –surface interaction force is

$$F_z = \left\langle \frac{\partial}{\partial z}\left(\mathbf{dE} + \mathbf{mB}\right) \right\rangle \tag{1}$$

where $\mathbf{d}$ and $\mathbf{m}$ are the particle fluctuation electric and magnetic dipole moments, $\mathbf{E}$ and $\mathbf{B}$ are the components of the fluctuating EMF. All these quantities have to be composed of spontaneous and induced terms (in the next section they are denoted by superscripts “sp” and “ind”). The angular brackets in (1) imply total quantum and statistical averaging. The space derivative is taken over coordinate $z$ of the rectangular system with the corresponding axis being directed along the surface normal (see Fig.1).

It should be noted that Eq. (1) can be adequately applied to both resting and moving particles embedded in fluctuating EMF, because it stems from the general expression for an average Lorentz force in the resting laboratory frame

$$\mathbf{F} = \int \langle \rho \mathbf{E} \rangle d^3 r + \frac{1}{c}\int \langle \mathbf{j} \times \mathbf{B} \rangle d^3 r \tag{2}$$

where $\rho$ and $\mathbf{j}$ are the fluctuating charge and current densities, and integrations are carried out over the particle volume. To demonstrate complete equivalence of (1), (2) and their relation with other expressions being currently used today [16], [17], let us consider an adiabatic motion of a particle near the surface with constant velocity $\mathbf{V}$. The corresponding vectors of electric and magnetic polarization originated by the particle are given by

$$\mathbf{P}(\mathbf{r},t)=\mathbf{d}(t)\delta(\mathbf{r}-\mathbf{V}t) \tag{3}$$

$$\mathbf{M}(\mathbf{r},t)=\mathbf{m}(t)\delta(\mathbf{r}-\mathbf{V}t) \tag{4}$$

Then, by substituting the charge and current densities $\rho=-div\mathbf{P}$, $\mathbf{j}=\frac{\partial\mathbf{P}}{\partial t}+c\cdot rot\mathbf{M}$ into (2) with account of (3), (4) and the Maxwell relations $rot\mathbf{E}=-\frac{1}{c}\frac{\partial\mathbf{B}}{\partial t}$, $div\mathbf{B}=0$, we directly obtain

$$\mathbf{F}=\nabla\left\langle\mathbf{dE}+\mathbf{mB}\right\rangle+\frac{1}{c}\left\langle\frac{\partial}{\partial t}(\mathbf{d}\times\mathbf{B})\right\rangle+\frac{1}{c}\left\langle(\mathbf{V}\nabla)(\mathbf{d}\times\mathbf{B})\right\rangle=\left\langle\nabla(\mathbf{dE}+\mathbf{mB})\right\rangle+\frac{1}{c}\left\langle\frac{d}{dt}(\mathbf{d}\times\mathbf{B})\right\rangle \tag{5}$$

Eq. (5) completely agrees with [16],[17]. For fluctuating $\mathbf{d}$ and $\mathbf{B}$ the last term in Eq. (5) becomes zero, and we are left with

$$\mathbf{F}=\left\langle\nabla(\mathbf{dE}+\mathbf{mB})\right\rangle \tag{6}$$

The same result $\frac{1}{c}\left\langle\frac{d}{dt}(\mathbf{d}\times\mathbf{B})\right\rangle=0$ was validated in Ref. [18] provided that $\mathbf{m}=0$. Eq. (6) is convenient one to calculating both attractive force to the surface, Eq. (1), and lateral, frictional force, as well [12],[13],[14],[15],[19]. In case of laser field or an externally applied constant EMF, Eq. (5) holds without statistical averaging and the last term should also be accounted for [16],[17].

To finish this preliminary discussion, we draw a special attention to the second term of Eq. (6), related with the magnetic moment $\mathbf{m}$. If the particle is moving, albeit its magnetic polarizability is negligibly small, the moment $\mathbf{m}$ in the laboratory frame appears due to the Lorentz transformations of the variables $\mathbf{P}$ and $\mathbf{M}$, thus giving rise to an additional contribution to the fluctuation electromagnetic force [12],[13],[14],[15],[19]. But if magnetic polarizability is not small, the magnetic coupling effect appears even for a resting particle. Just this issue and the involved dependences of the resulting force on distance, temperature and material properties will be properly address in this letter.

## 2.Theory

Despite subsequent calculation in the right hand side of (1) represents a rather tedious technical problem, it turns out to be successfully worked out by following the previously developed scheme, step by step [12],[13],[14],[15]. According to this one, the induced field components, $\mathbf{E}^{\mathrm{ind}},\mathbf{B}^{\mathrm{ind}}$ can be directly obtained solving the Maxwell equations with extraneous fluctuating currents, induced by moments $\mathbf{d}^{\mathrm{sp}},\mathbf{m}^{\mathrm{sp}}$ (see (3),(4) and the relations below), while all the quantities have to be represented as Fourier transforms over the time and space variables $t,x,y$. The involved correlators of dipole moments are simplified using the fluctuation –dissipation theorem [20]. On the other hand, the induced components $\mathbf{d}^{\mathrm{ind}},\mathbf{m}^{\mathrm{ind}}$ have to be expressed through the fluctuating fields $\mathbf{E}^{\mathrm{sp}},\mathbf{B}^{\mathrm{sp}}$ of the surface and the particle electric/magnetic polarizabilities $\alpha_{e,m}(\omega)$, by using linear integral relations. The corresponding correlators of fluctuating EMF are calculated via components of the retarded photon Green function [20]. Finally, assuming the same temperature $T$ both for the particle and surface, the attraction force is found to be

$$\begin{aligned} F_z = &-\frac{2\hbar}{\pi^2}\int_0^\infty d\omega \coth\left(\hbar\omega/2k_BT\right)\int_{k>\omega/c} d^2k \exp\left(-2q_0z\right)\operatorname{Im} Z(\omega,k) - \\ &-\frac{2\hbar}{\pi^2}\int_0^\infty d\omega \coth\left(\hbar\omega/2k_BT\right)\int_{k<\omega/c} d^2k \operatorname{Im}\left(\exp\left(2\mathrm{i}\tilde{q}_0z\right)\tilde{Z}(\omega,k)\right) \end{aligned} \tag{7}$$

$$\begin{aligned} Z(\omega,k) = &\alpha_e(\omega)\left[\left(2k^2-\frac{\omega^2}{c^2}\right)\Delta_e(\omega,k)+\frac{\omega^2}{c^2}\Delta_m(\omega,k)\right]+ \\ &+\alpha_m(\omega)\left[\left(2k^2-\frac{\omega^2}{c^2}\right)\Delta_m(\omega,k)+\frac{\omega^2}{c^2}\Delta_e(\omega,k)\right] \end{aligned} \tag{8}$$

where $\hbar, k_B$ are the Planck's and Boltzmann's constants, $z$ is the particle –surface separation, $c$ – the speed of light in vacuum, $q_0=(k^2-\omega^2/c^2)^{0.5}$, $\tilde{q}_0=(\omega^2/c^2-k^2)^{0.5}$, the function $\tilde{Z}(\omega,k)$ is given by Eq. (8) with the replacement $\Delta_{e,m}(\omega,k)\to\tilde{\Delta}_{e,m}(\omega,k)$, whereas the proper surface response functions are written by

$$\Delta_e = \frac{\varepsilon\sqrt{k^2-\omega^2/c^2}-\sqrt{k^2-\varepsilon\mu\omega^2/c^2}}{\varepsilon\sqrt{k^2-\omega^2/c^2}+\sqrt{k^2-\varepsilon\mu\omega^2/c^2}} \quad \Delta_m = \frac{\mu\sqrt{k^2-\omega^2/c^2}-\sqrt{k^2-\varepsilon\mu\omega^2/c^2}}{\mu\sqrt{k^2-\omega^2/c^2}+\sqrt{k^2-\varepsilon\mu\omega^2/c^2}} \tag{9}$$

$$\tilde{\Delta}_e = \frac{\varepsilon\sqrt{\omega^2/c^2-k^2}-\sqrt{\varepsilon\mu\omega^2/c^2-k^2}}{\varepsilon\sqrt{\omega^2/c^2-k^2}+\sqrt{\varepsilon\mu\omega^2/c^2-k^2}}, \quad \tilde{\Delta}_m = \frac{\mu\sqrt{\omega^2/c^2-k^2}-\sqrt{\varepsilon\mu\omega^2/c^2-k^2}}{\mu\sqrt{\omega^2/c^2-k^2}+\sqrt{\varepsilon\mu\omega^2/c^2-k^2}} \tag{10}$$

Moreover, $\varepsilon$ and $\mu$ are the frequency –dependent dielectric and magnetic permittivities of the surface material. In (9), (10) all the arguments of the functions $\varepsilon,\mu$ and $\Delta_{e,m},\tilde{\Delta}_{e,m}$ are omitted for brevity.

The first integral in (7) describes particle –surface coupling via evanescent modes of EMF, the second one –via radiation modes. One sees that Eqs. (7), (8) manifest complete transposition symmetry over the electric (marked by subscript "e") and magnetic (marked by subscript "m") quantities. In the simplest case $\alpha_m(\omega)=0, T=0$ Eq. (7) reduces to

$$F_z=-\frac{\hbar}{\pi}\int_0^\infty d\omega\int_0^\infty dkk\exp\left(-2\sqrt{k^2+\omega^2/c^2}\,z\right)\cdot\alpha_e(\mathrm{i}\omega)\left[(2k^2+\omega^2/c^2)\Delta_e(\mathrm{i}\omega)-\frac{\omega^2}{c^2}\Delta_m(\mathrm{i}\omega)\right] \qquad (11)$$

Moreover, the expression for the van –der –Waals (nonretarded) force stems from (11), (9) in the limit $c\to\infty$:

$$F_z^{(vdW)}=-\frac{3}{4\pi}\frac{\hbar}{z^4}\int_0^\infty\frac{\varepsilon(\mathrm{i}\omega)-1}{\varepsilon(\mathrm{i}\omega)+1}\alpha_e(\mathrm{i}\omega)d\omega \qquad (12)$$

In order to study the general case, let us rewrite Eq. (7) in the form $F_z=F_z^{(1)}+F_z^{(2)}$, where the first term is assumed to be temperature independent (except for the material properties involved). This can be done by using the identity $\coth(x/2)=1+2/(\exp(x)-1)$. Then, after making use of trivial transformations related with the frequency and wave –vector variables, we get from (7)

$$F_z^{(1)}=-\frac{\hbar}{\pi}\int_0^\infty d\omega\left(\frac{\omega}{c}\right)^4\int_1^\infty duu\,\exp\left(-2\omega zu/c\right)\cdot$$
$$\cdot\left\{\begin{array}{l}\left[\left(2u^2-1\right)\tilde{\Delta}_e(u,\varepsilon)-\tilde{\Delta}_m(u,\varepsilon)\right]\alpha_e(\mathrm{i}\omega)+\\+\left[\left(2u^2-1\right)\tilde{\Delta}_m(u,\varepsilon)-\tilde{\Delta}_e(u,\varepsilon)\right]\alpha_m(\mathrm{i}\omega)\end{array}\right\},\ \varepsilon=\varepsilon(\mathrm{i}\omega) \qquad (13)$$

$$F_z^{(2)}=-\frac{2\hbar}{\pi}\int_0^\infty d\omega\left(\frac{\omega}{c}\right)^4\left[\exp\left(\hbar\omega/kT\right)-1\right]^{-1}\cdot$$
$$\cdot\left\{\begin{array}{l}\int_0^1 duu\ \mathrm{Im}\left[\begin{array}{l}\exp\left(2\mathrm{i}\omega zu/c\right)\left(\left(1-2u^2\right)\tilde{\Delta}_e(u,\varepsilon)+\tilde{\Delta}_m(u,\varepsilon)\right)\alpha_e(\omega)+\\+\exp\left(2\mathrm{i}\omega zu/c\right)\left(\left(1-2u^2\right)\tilde{\Delta}_m(u,\varepsilon)+\tilde{\Delta}_e(u,\varepsilon)\right)\alpha_m(\omega)\end{array}\right]+\\+\int_0^\infty duu\exp\left(-2\omega zu/c\right)\mathrm{Im}\left[\begin{array}{l}\left(\left(2u^2+1\right)\Delta_e(u,\varepsilon)+\Delta_m(u,\varepsilon)\right)\alpha_e(\omega)+\\+\left(\left(2u^2+1\right)\Delta_m(u,\varepsilon)+\Delta_e(u,\varepsilon)\right)\alpha_m(\omega)\end{array}\right]\end{array}\right\} \qquad (14)$$

$$\Delta_e=\frac{\varepsilon u-\sqrt{u^2+1-\varepsilon\mu}}{\varepsilon u+\sqrt{u^2+1-\varepsilon\mu}},\ \Delta_m=\frac{\mu u-\sqrt{u^2+1-\varepsilon\mu}}{\mu u+\sqrt{u^2+1-\varepsilon\mu}} \qquad (15)$$

$$\tilde{\Delta}_e=\frac{\varepsilon u-\sqrt{u^2+\varepsilon\mu-1}}{\varepsilon u+\sqrt{u^2+\varepsilon\mu-1}},\ \tilde{\Delta}_m=\frac{\mu u-\sqrt{u^2+\varepsilon\mu-1}}{\mu u+\sqrt{u^2+\varepsilon\mu-1}} \qquad (16)$$

In writing (13), (14) we took $\mu=1$ and thus involved arguments of the response functions $\Delta_{e,m},\tilde{\Delta}_{e,m}$ were omitted. Note also, that in (14) the function $\varepsilon(\omega)$ is computed at real frequencies $\omega$. It is worthwhile noting that Eqs.(12),(13) exactly coincide with the corresponding formulae obtained in Ref.[21] at $\alpha_m(\omega)=0$, but, finally, Eqs.(7), (13) and (14) manifest principally new consequences.

To get meaningful quantitative results, we use Eqs. (13), (14) to calculating the retarded force between a spherical metallic nanoparticle of radius $R$ and the metallic wall. For simplicity, both materials are assumed to be the same normal nonmagnetic metals ($\mu=1$) having the Drude –like bulk dielectric functions

$$\varepsilon(\omega)=1-\frac{\omega_p{}^2}{\omega(\omega+\mathrm{i}/\tau)} \quad , \tag{17}$$

where $\omega_p$ and $\tau$ are the plasma frequency and the relaxation time. Using classical approximation [1], the functions $\alpha_{e,m}(\omega)$ are given by

$$\alpha_e(\omega)=R^3\frac{\varepsilon(\omega)-1}{\varepsilon(\omega)+2}\equiv R^3\varphi_e(\omega)\,, \tag{18}$$

$$\alpha_m(\omega)=-\frac{R^3}{2}\left[1-\frac{3}{R^2k^2}+\frac{3}{Rk}\cot(Rk)\right]\equiv-\frac{R^3}{2}\varphi_m(\omega), k=(1+\mathrm{i})\sqrt{2\pi\sigma\omega}/c \tag{19}$$

where $\sigma=\sigma_0/(1-\mathrm{i}\omega\tau)$, $\sigma_0$ is the static conductivity. Moreover, at an imaginary frequency $\mathrm{i}\omega$ the function $\alpha_m(\omega)$ takes the form

$$\alpha_m(\mathrm{i}\omega)=-\frac{R^3}{2}\left[1+\frac{3}{y^2}-\frac{3}{y}\coth y\right]\equiv-R^3\tilde{\varphi}_m(y), y=\frac{2R}{c}\sqrt{\frac{\pi\sigma_0\omega}{1+\omega\tau}} \tag{20}$$

By introducing the reduced frequency variable $x=2\omega zu/c$, denoting $\lambda_P=2\omega_P z/c, \lambda_\tau=2z/\tau c, \lambda_R=2\pi\sigma_0 R/c$ and taking account of (17)-(20), Eq. (13) is transformed to

$$F_z^{(1)}=-\frac{\hbar c}{32\pi}\frac{R^3}{z^5}\int_0^\infty dx x^4 e^{-x}\int_1^\infty\frac{du}{u^4}\left\{\tilde{R}_e(u,x)\tilde{\varphi}_e(cx/2zu)+\tilde{R}_m(u,x)\tilde{\varphi}_m\left(y(x/u)\right)\right\} \tag{21}$$

$$\tilde{R}_e(u,x)=(2u^2-1)\tilde{\Delta}_e(u,\varepsilon)-\tilde{\Delta}_m(u,\varepsilon) \tag{22}$$

$$\tilde{R}_m(u,x)=(2u^2-1)\tilde{\Delta}_m(u,\varepsilon)-\tilde{\Delta}_e(u,\varepsilon) \tag{23}$$

$$y(x/u)=\frac{(R/z)^{0.5}\lambda_R{}^{0.5}(x/u)^{0.5}}{(1+\lambda_\tau x/u)^{0.5}} \tag{24}$$

$$\varepsilon=1+\frac{\lambda_p{}^2u^2}{x^2(1+xu\lambda_\tau)} \tag{25}$$

$$\tilde{\varphi}_e(cx/2zu)=\frac{\varepsilon(\mathrm{i}\omega)\text{-}1}{\varepsilon(\mathrm{i}\omega)\text{+}2}=\frac{\lambda_p{}^2u^2}{3x^2(1+xu\lambda_\tau)+\lambda_p{}^2u^2} \tag{26}$$

Furthermore, introducing $x=\omega/\omega_W$, $\omega_W=k_B T/\hbar$, $\lambda_W=\frac{2\omega_W z}{c}$, $\lambda_{0W}=\frac{2R}{c}(\pi\sigma_0\omega_W)^{0.5}$, the thermal part of the interaction force $F_z^{(2)}$ is given by

$$F_z^{(2)}=F_z^{(2,ev)}+F_z^{(2,rad)}=$$
$$-\frac{\hbar\omega_W}{8\pi}\frac{R^3}{z^4}\int_0^\infty\frac{dx}{e^x-1}\int_0^\infty dyy\exp(-y)\,\mathrm{Im}\left[R_e(y/x\lambda_W,\varepsilon)\varphi_e(\omega_W x)+R_m(y/x\lambda_W,\varepsilon)\varphi_m(\tilde{x})\right]-$$
$$+\frac{2\hbar\omega_W}{\pi}\left(\frac{\omega_W R}{c}\right)^3\frac{\omega_W}{c}\int_0^\infty\frac{dxx^4}{e^x-1}\int_0^1 duu\,\mathrm{Im}\left[\exp(\mathrm{i}\lambda_W xu)\begin{pmatrix}\tilde{R}_e(u,\varepsilon)\varphi_e(\omega_W x)+\\+\tilde{R}_m(u,\varepsilon)\varphi_m(\tilde{x})\end{pmatrix}\right] \quad (27)$$

$$R_e(y/x\lambda_W,\varepsilon)=(2y^2+\lambda_W{}^2)\Delta_e(y/x\lambda_W,\varepsilon)+\lambda_W{}^2\Delta_m(y/x\lambda_W,\varepsilon) \quad (28)$$

$$R_m(y/x\lambda_W,\varepsilon)=(2y^2+\lambda_W{}^2)\Delta_m(y/x\lambda_W,\varepsilon)+\lambda_W{}^2\Delta_e(y/x\lambda_W,\varepsilon) \quad (29)$$

$$\tilde{x}=\lambda_{0W}\left(\frac{-x}{\omega_W\tau x+\mathrm{i}}\right)^{0.5} \quad (30)$$

where $\Delta_{e,m}(u,\varepsilon)$ is evaluated from (15) at $\mu=1$ and $\varepsilon$ is given by (17) at $\omega=x\omega_W$.

## 3.General comments and numerical results

A general analysis of the expressions (21) and (27) shows the leading force -distance dependences on $z$ to be : $F_z^{(1)}\propto z^{-5}$ in the case of “cold” interaction force, Eq.(21), and $F_z^{(2,ev)}\propto z^{-4}$ in the case of evanescent “thermal” force, Eq.(27). The “radiation” thermal contribution, $F_z^{(2,rad)}$(the second term (27)), as it stems from the structure of the proper inner integral, reveals an oscillating behavior. The additional (much weaker) $z-$dependences of the integrals (21), (27) come through the reduced parameters $\lambda_p,\lambda_\tau$ and $\lambda_W$.

Because $F_z^{(2,rad)}\sim\left(\frac{\omega_W R}{c}\right)^3$ (Eq.(27)), this force becomes appreciable only for relatively big particles, if $\omega_W R/c\approx 1$. At normal temperature, $T=300K$ this implies $R\approx 7.7\mu m$, however, this limit decreases with increasing temperature. The main temperature dependence of $F_z^{(2,ev)}$ is $\sim T$, whereas for $F_z^{(2,rad)}$ it turns out to be much stronger $\sim T^5$ , and therefore the radiation force turns out to be dominating for micron size particles with increasing temperature.

As a numerical example, we have considered vacuum contact between a Cu particle and a Cu substrate. In this case at $T$=300 $K$ we have $\sigma_0=5.2\cdot 10^{17}s^{-1}$, $\tau=2.5\cdot 10^{-14}s$ and $\omega_p = \sqrt{4\pi\sigma_0 / \tau} = 1.62 \cdot 10^{16}\, s^{-1}$ [23]. First, we have computed the "cold" force $F_z^{(1)}$ both with and without magnetic polarization term, which is given by the second addend in Eq. (21). The fractions of the calculated contributions *versus* radius $R$ at $R/z = 1/3$ and $R/z = 1/10$ are shown in Fig. 2. Hereafter we have also taken into account correct temperature dependences for $\sigma_0$ and $\tau$ [23], and thus the solid and dotted lines in Fig.2 show the obtained results at $T = 300K$, while the next two lines (dash and dash-dotted lines) correspond to $T = 77K$. We see that magnetic polarization gives rise to larger interaction force: by 1.5 times as maximum for the component $F_z^{(1)}$. However, this growth is visible only for particles of radii larger than several tens *nm*.

Second, we have analyzed relative contribution of the temperature effect on total force. Fig.3 shows the fraction $F_z(T)/F_z(0)$ vs. distance *z* at $R = 1\mu m$ and at different temperature *T*. For particles of smaller radii ($R < 1\mu m$) the calculations show that the temperature effect subsides and the results reveal usual attractive behavior of the retarded Casimir force, which is described by the component $F_z^{(1)}$.

As one can see from Fig.3, the obtained temperature dependence is significant. Moreover, negative sign of the line labeled "$600K$" at $z > 12\mu m$ signifies onset of repulsive interaction. The repulsion force becomes still more evident for bigger particles with increasing temperature. Lines 1 to 3 in Figs. 4, 5 show the total interaction forces (Eqs.(21) and (27)) for a particle of radius $R = 3\mu m$ at $T = 77, 300$ and $600K$ (Fig.4) and of radius $R = 4\mu m$ at $T = 300, 600, 900K$ (Fig.5). Line 4 in Fig.5 corresponds to $T = 600K$ without magnetic polarization term. Note an opposite sign of the magnetic coupling effect (cf. lines 2 and 4 in Fig.5) in comparison with Fig.2, where the magnetic polarization is seen to enlarge modulus of attraction force. Also, it is worthwhile noting that both in the cases of Fig.4 and Fig.5 the dominating contributions to the interaction stem from the component $F_z^{(2,rad)}$, Eq. (27). By analogy with Ref. [10], one may call this force "repulsive wind". Note that authors of [10] calculated interaction force between a small SiC particle above a heated SiC substrate in the very near field. In that case, however, the magnetic coupling was negligibly small, while the near field of radiating substrate and particle was characterized by neighboring resonance frequencies. Moreover, in our calculations both the particle and surface have the same temperature $T$, whereas in [10] the particle was assumed to be held at zero temperature.

## 4.Summary

To sum up, we conclude that fluctuation magnetic polarization of small metallic particle being placed nearby a metallic surface, of nonmagnetic materials, gives rise to an important contribution to the fluctuation electromagnetic force, both at nanoscale separations, when the “cold” part of the interaction force dominates (due to quantum EMF fluctuations), and at microscale (for bigger particles), when the interaction force proves to be dominated by thermal EMF fluctuations. For the first time, using dipole approximation, we have obtained general theoretical expressions for the corresponding forces which take account of electric and magnetic polarization terms, thermal and retardation effects. The numerical calculations show that the resulting force manifests principally new dependences on temperature, particle dimension, separation and material properties. Particularly, the long range interaction force for micron size metallic particle with metallic surface turns out to be repulsive. It is originated by thermal electromagnetic fluctuations (radiation wind), weakly depends on distance to the surface and noticeably exceeds the classical Casimir force at large enough temperature.

**Acknowledgments**

This work in part was supported by Russian Foundation for Basic Research, the grant No. 06-02-17140.

**Figure captions:**

Fig.2 Fraction of the forces $F_z^{(1)}$ vs. radius $R$ calculated both with and without magnetic polarization term ($\alpha_m(\omega)=0$ ) between a Cu particle and a Cu substrate. Solid line - $R/z=1/3, T=300K$, dotted line - $R/z=1/10, T=300K$, dashed line - $R/z=1/3, T=77K$, dash –dotted line - $R/z=1/10, T=77K$. The temperature dependence of material properties is properly accounted for.

Fig.3 Fraction of the forces $F_z(T)/F_z^{(1)}(0)$ *vz.* distance *z* for Cu particle of $R=1\mu m$ above Cu surface. Lines 1 and 2 correspond to $T=300, 600$, respectively.

Fig.4 Dependence $F_z(T)$ vs. distance *z* at $R=3\mu m$. Lines 1 to 3 correspond to $T=77, 300$ and 600*K.* The temperature dependence of material properties is taken into account.

Fig.5. The same as in Fig.4 at $R=4\mu m$. Lines 1 to 3 correspond to $T=300{,}600{,}900K$. Line 4 has been calculated without magnetic polarization term at $T=300K$.

Fig.1

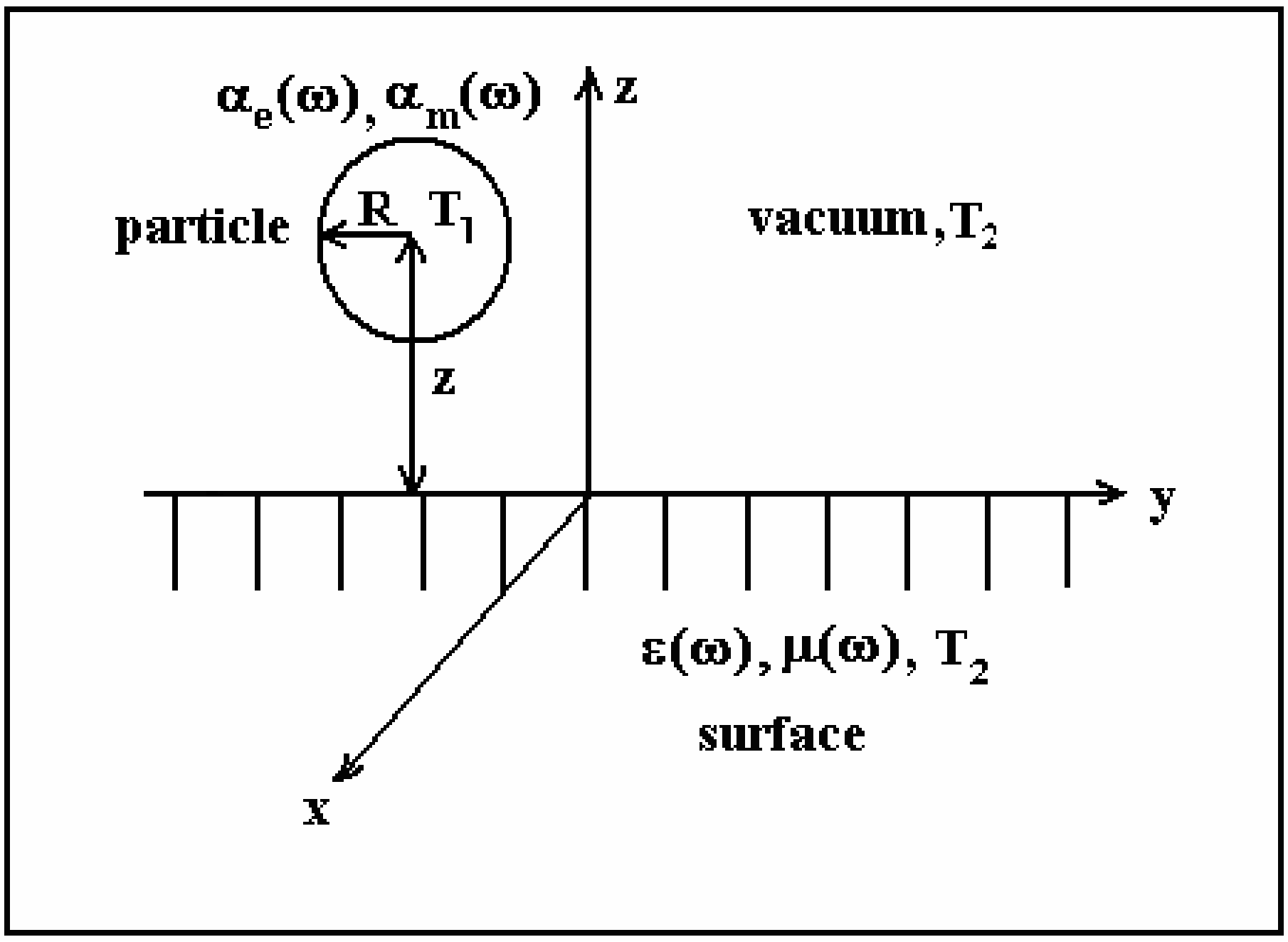

Fig.2

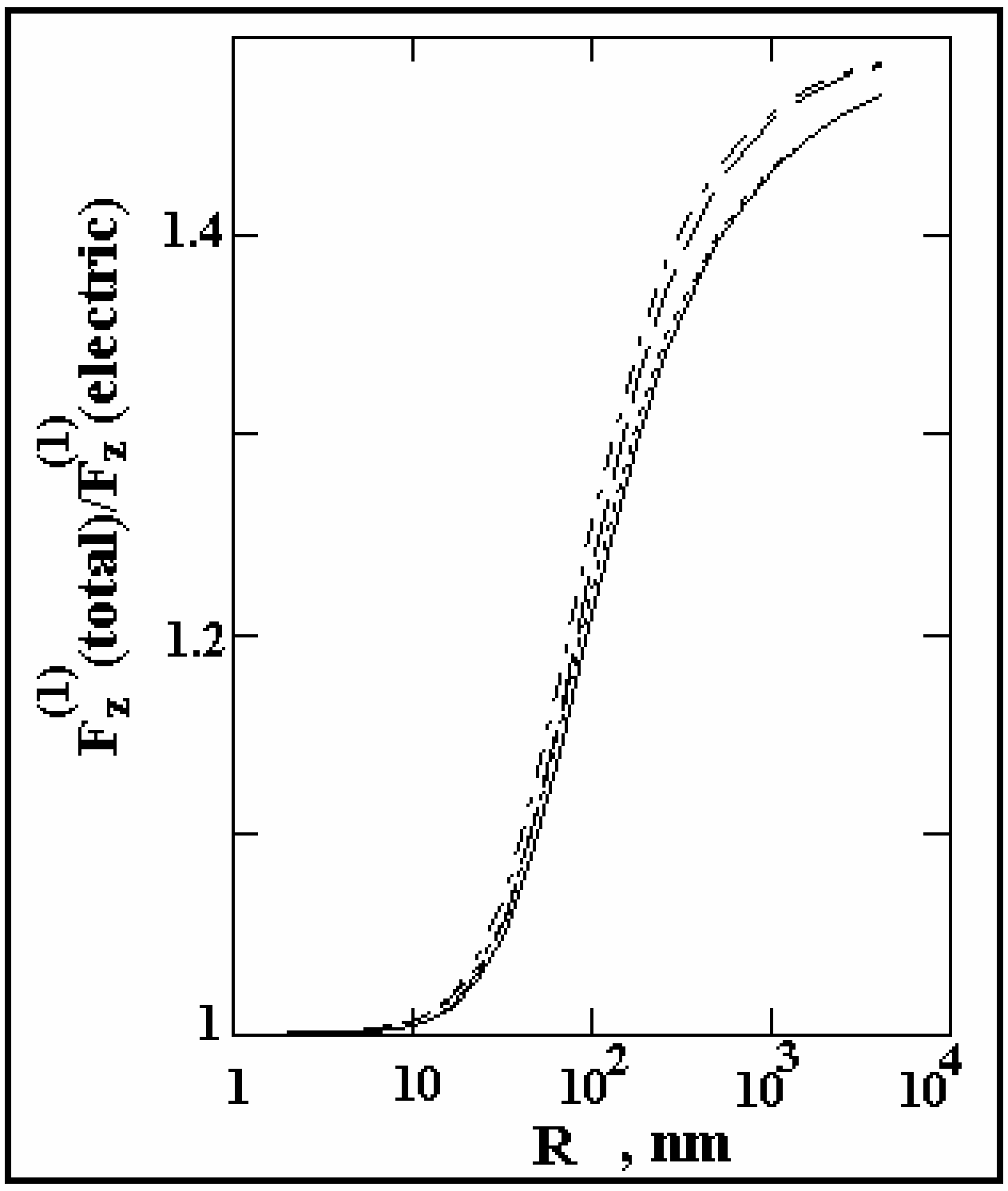

Fig.3

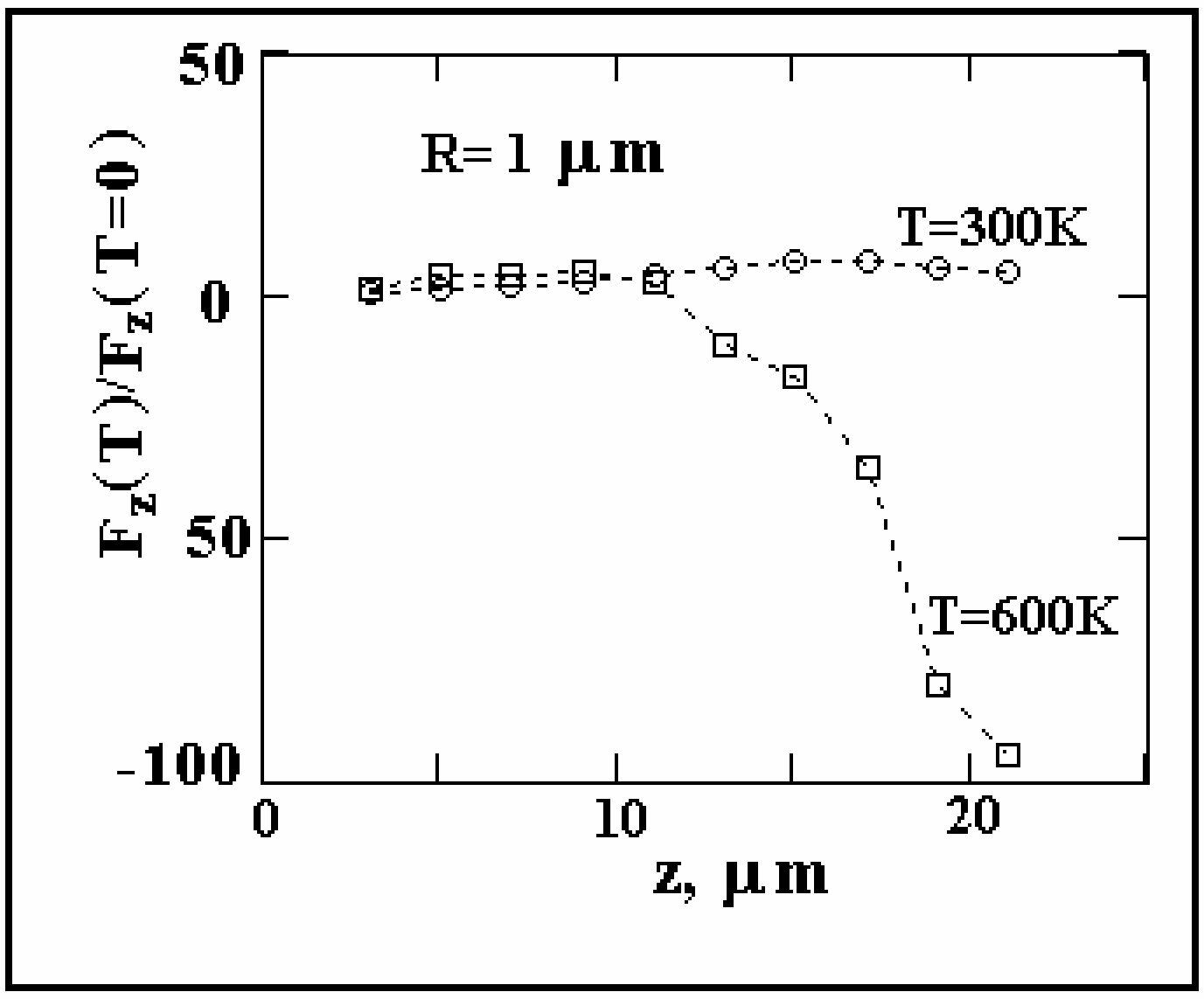

Fig.4

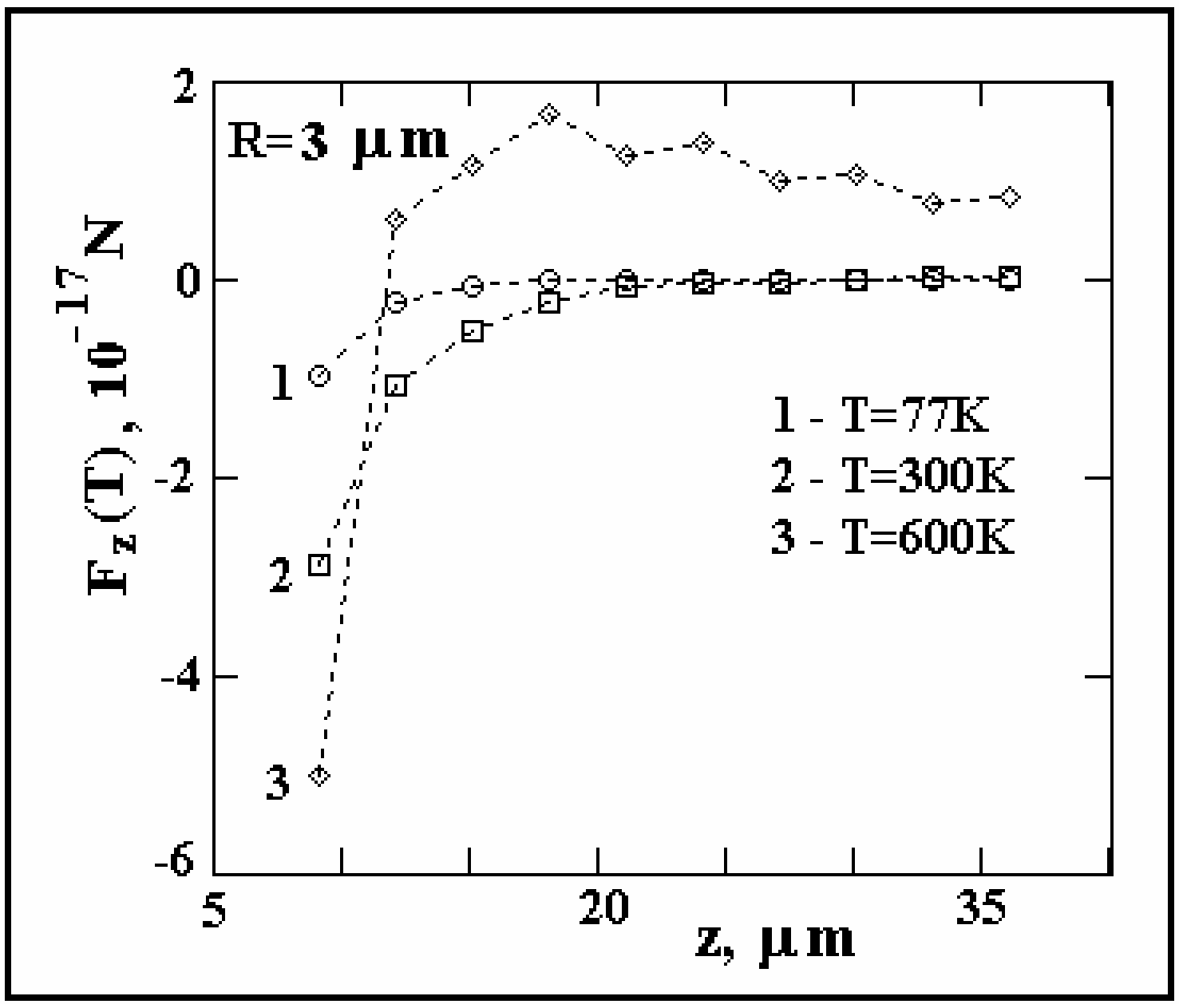

Fig.5

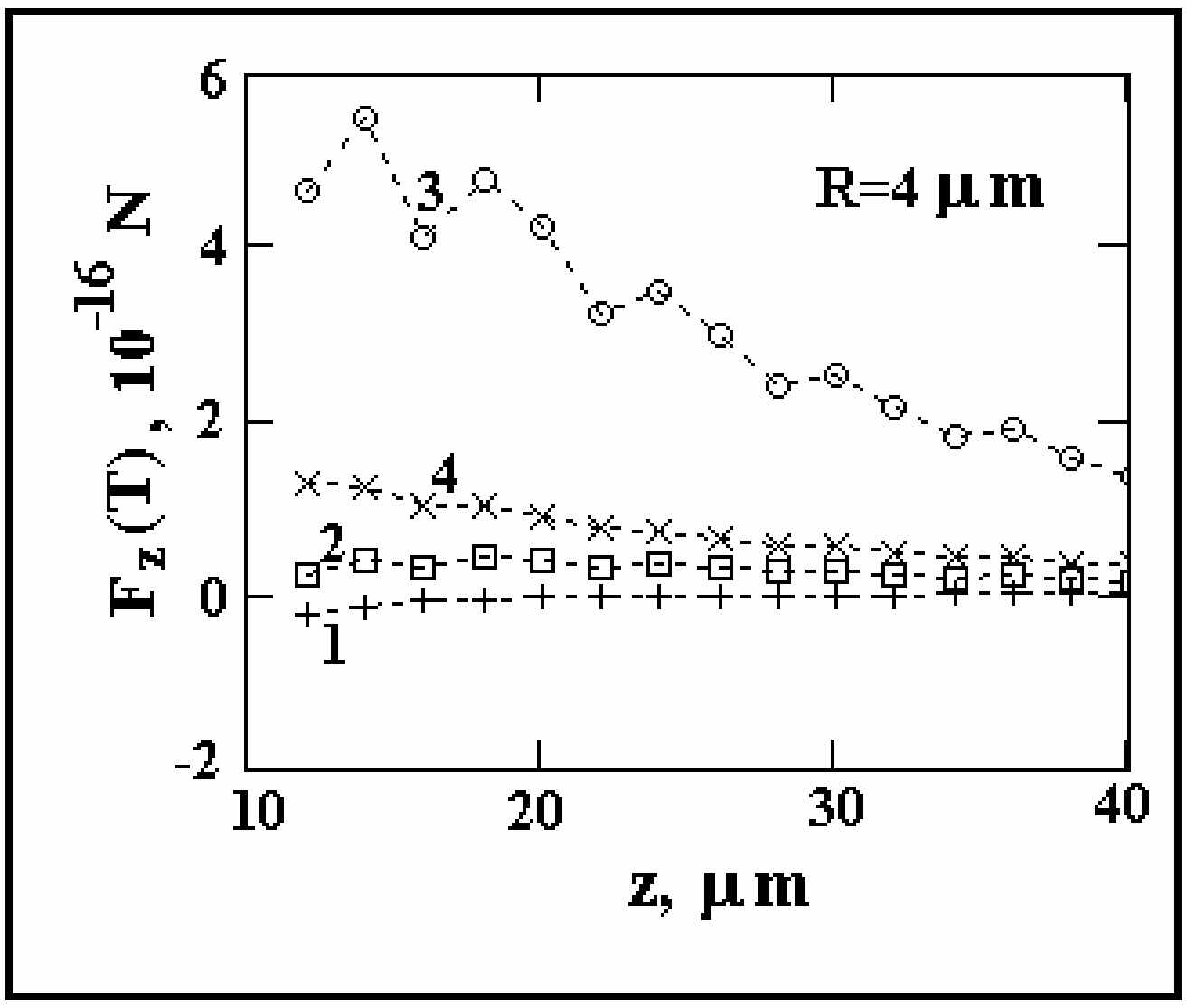